\documentclass[twocolumn,secnumarabic,amssymb, floatfix, nobibnotes, aps,superscriptaddress]{revtex4-1}
\usepackage{hyperref}
\pdfoutput=1

\usepackage{float}
\usepackage{graphicx}
\usepackage{amsmath} 
\usepackage{epstopdf}
\usepackage{setspace}
\usepackage[shortlabels]{enumitem}
\usepackage{mathrsfs}
\usepackage{color,soul}
\usepackage[dvipsnames]{xcolor}
\usepackage{dcolumn,amssymb}
\usepackage{comment}
\usepackage[normalem]{ulem}
\usepackage{stackengine}
\usepackage{amsmath}

\usepackage{scalerel}

\stackMath
\def\dydot{.2ex}
\newcommand\caldot[1]{\ThisStyle{%
              \stackon[\dydot]{\SavedStyle\mathcal{#1}}
                              {\SavedStyle\dot{\phantom{\mathcal{#1}}}}}}
\usepackage{dcolumn} 
\usepackage{natbib}
\usepackage{bm} 
\usepackage{float} 
\usepackage{bbm} 
\usepackage{color}
\usepackage{amsfonts}
\usepackage{lmodern} 
\usepackage{amsmath,amssymb}
\hypersetup{colorlinks=true,citecolor=blue,linkcolor=red,urlcolor=blue}

\renewcommand{\vec}[1]{\boldsymbol{#1}}
\newcommand{\tens}[1]{\boldsymbol{#1}}

\begin{document}
\title{Coarse-graining dense, deformable active particles}
\author{Mehrana R. Nejad}
\affiliation{Department of Physics, Harvard University, Cambridge, MA 02138}
\affiliation{School of Engineering and Applied Sciences, Harvard University, Cambridge, MA 02138, USA}
\email{mehrana@g.harvard.edu}
\author{Julia M. Yeomans}
\affiliation{The Rudolf Peierls Centre for Theoretical Physics, Department of Physics, University of Oxford, Parks Road, Oxford OX1 3PU, United Kingdom}
\begin{abstract}
We coarse-grain a model of closely-packed ellipses that can vary their aspect ratio to derive continuum equations for materials comprising confluent deformable particles such as epithelial cell layers. We show that contractile nearest neighbour interactions between ellipses can lead to their elongation  and nematic ordering. Adding flows resulting from active hydrodynamic stresses produced by the particles also affects the aspect ratio and can result in active turbulence. Our results, which agree well with multi-phase field simulations of deformable isotropic cells,  provide a bridge between models which explicitly resolve cells and continuum theories of active matter.
\end{abstract}
\maketitle

\begin{figure*}[t] 
    \centering
\includegraphics[width=0.95\textwidth]{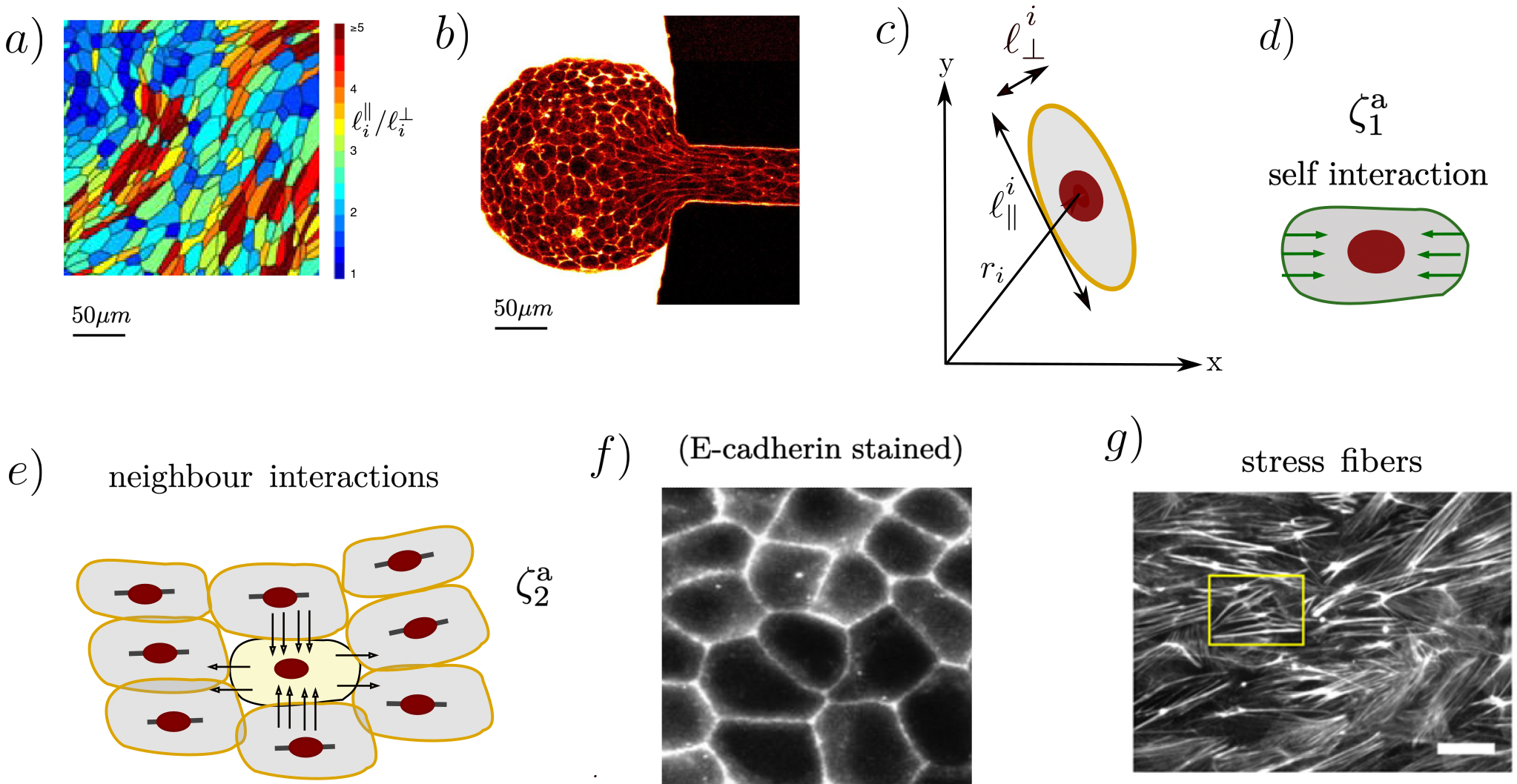}
    \caption{(a) Activity leads to cell elongation in epithelial layers. The colour code shows the aspect ratio of the cells (adapted from \cite{mitchel2020primary}). (b) Elongation of cells in an aggregate forced to flow through a constriction in a microfluidic channel  \cite{tlili2022microfluidic}.  Schematic of (c) the anisotropic Gaussian field, Eq.~(\ref{eq3}); (d) the self-interaction, Eq.~(\ref{eq5}), (e) the neighbour interaction, Eq.~(\ref{eq6}),
   resulting from (f) cell-cell junctions and (g) stress fibres exerting forces on the surrounding cells \cite{eckert2023hexanematic,balasubramaniam2021investigating,choi2022cell,haas2022zo}.
    The yellow box highlights the approximate size of a cell. (f) and (g) are adapted from \cite{le1999recycling} and \cite{kemp2018actin}, respectively.} 
  \label{fig1main}
\end{figure*}

\noindent
 There are many instances where the way particles can deform is important in controlling their behaviour. For example, eukaryotic cells continually change shape as they move through the extra-cellular matrix \cite{fletcher2010cell,geiger2009environmental} and the shape of colloidal microgels can be controlled by temperature or humidity \cite{tan2008review,lefroy2021advances} and flows \cite{PhysRevLett.96.028104,dupire2012full,abkarian2007swinging,PhysRevE.107.044608}. Particle shape changes can also affect the collective properties of active and soft matter. Active forces result in elongation and local nematic order in epithelial monolayers, which in turn allows active turbulent flows and the creation of topological defects \cite{atia2018geometric,PhysRevLett.121.248003,nejad24}, and convergent-extension of tissues is a common step in embryogenesis  (see also Fig.~\ref{fig1main}a,b). The flow properties of emulsions and foams is closely related to how they can deform under shear \cite{cantat2013foams,exerowa2018foam}.

Methods that resolve individual cells, such as multi-phase field and vertex models, can naturally include variations in the shape and the connectivity of individual particles. However current continuum approaches to modelling confluent cell layers and tissues rely on active nematohydrodynamics which assumes that the shape of the constituent nematogens remains fixed \cite{zhang2022topological,PhysRevX.12.041017,saw2017topological,ascione2022collective}. yt6It is not clear how to include the shape changes of individual particles in the theories of active (or passive) nematohydrodynamics, or the extent to which this will change the physics. As a step in this direction, we coarse-grain a model of dense, deformable ellipses to construct a continuum description of deformable nematogens. We show how shape-changing, short-range interactions can be incorporated in a manner analogous to including a Landau de Gennes free energy in fixed-particle-shape nematohydrodynamic models, and argue how to include long-range active hydrodynamic flows.
 
We consider a confluent system and work in terms of 
 the coarse-grained variables ${\mathbf{W}}$ and $\mathcal{A}$.  $\mathbf{W}$ is the anisotropy tensor which, in two dimensions, is a $2 \times 2$ symmetric, traceless tensor that describes the coarse-grained magnitude and direction of particle {\em elongation}. (This is in contrast to standard nematohydrodynamics theories which consider the average {\em orientation} of elongated particles using a nematic tensor, $\mathcal{Q}$ \cite{duclos2018spontaneous,lavi2024nonlinear,Santhosh}.) $\mathcal{A}$ is a variable describing the locally averaged particle area. 

We assume that each of the hydrodynamic variables $\mathcal{X} =\{\mathbf{W},\mathcal{A}\}$ obeys an equation of motion of the form
\begin{equation}
\partial_t \mathcal{X} =  \caldot X^f +  \caldot X^I  + \caldot X^u
\label{eqmotion}
\end{equation}
where $\caldot X^I$  are terms from short-range, dipolar interactions that can change particle shape. We first derive these terms, considering both forces internal to each nematogen and nearest-neighbour interactions between the particles.
We then discuss $ \caldot X^f $,
which is similar to the fixed-particle-shape interactions that are commonly included in liquid crystal hydrodynamics as a Landau expansion in the order parameters. We show how these terms contributions with the dipolar forces to control the elongation and nematic ordering of the particles and compare the results to mesoscale simulations of confluent cell layers based on multi-phase field models \cite{PhysRevLett.130.038202,chiang2023intercellular}. Finally, we incorporate the response of the nematogens to a flow field $\vec{u}$, $\caldot X^u$, and we discuss the active turbulence that results from a flow field driven by active hydrodynamic stresses. \\

\noindent
\textit{Active short-range interactions}—
\label{activeshortrange}
We model individual active particles as deformable ellipses represented by Gaussian fields. The Gaussian representation is advantageous because it allows analytical progress in calculating the active particle dynamics.  Particle $i$ at position $\textbf{r}_i$, with orientation $\theta_i$, director $\vec{m}_{i}= (\cos{\theta_i},\sin{\theta_i})$,  and long and short axis $l^{\parallel}_i$ and $l_i^{\perp}$, is described by the field
\begin{equation}
\varphi_i(\textbf{r}) = e^{- \left(\frac{\textbf{r}-\textbf{r}_i}{2}\right) \cdot  \left( \frac{\textbf{I}+\textbf{q}_i}{(\ell_i^{\parallel})^2} + \frac{\textbf{I}-\textbf{q}_i}{( \ell_i^{\perp})^2}  \right)\cdot \left(\frac{\textbf{r}-\textbf{r}_i}{2}\right)},\label{eq3}
\end{equation}
where $\textbf{q}_i= (2 \vec{m}_{i} \vec{m}_{i} - \textbf{I})$ is the orientation tensor with $(q_i)_{11}=-(q_i)_{22}=\cos{2 \theta_i}$, and $(q_i)_{12}^i=(q_i)_{21}=\sin{2 \theta_i}$.
Fig.~\ref{fig1main}(c) shows a schematic of an ellipse.  We will work to linear order in the particle anisotropy and area change and derive equations of motion for the dynamics of the shape tensor 
\begin{align}
\label{eq:deformation_tensor}
\boldsymbol{{d}}_i
&= - \int d\mathbf{x} \left \{ \boldsymbol{\nabla} \varphi_i \boldsymbol{\nabla} \varphi_i^T - {\frac{\mathbf{I}}{2}} \mbox{Tr}( \boldsymbol{\nabla} \varphi_{i} \boldsymbol{\nabla} \varphi_i^T )\right\}\nonumber\\ &= \frac{\pi^2}{4 a_i} \textbf{w}_i, 
\end{align} 
where we have defined cell area $a_i=\pi \ell_i^{\parallel} \ell_i^{\perp}$, and cell anisotropy tensor $\textbf{w}_i=\{(\ell^{\parallel}_{i})^2-(\ell^{\perp}_i)^2 \}\textbf{q}_i$.

Guided by recent work on confluent cell layers, and observation of nematic $\pm1/2$ defects, we assume that the short-range forces acting on the ellipses have nematic symmetry \cite{armengol2024hydrodynamics,saw2017topological,nejad24}. We will use the terminology {\em active} for these forces because they are driven by biological processes in individual cells. 
Using the phase field introduced in Eq.~(\ref{eq3}) the internal active nematic stress for the particle $i$ can be written
\begin{equation}
\boldsymbol{\pi}^{s}_i= -\zeta_1^\text{a} \mathbf{d}_i {\varphi_i}(\mathbf{x}) ,\label{eq5}
\end{equation}
where $\zeta_1^\text{a}$ is the magnitude of the stress. We will refer to this term as the self-interaction. We consider $\zeta_1^\text{a}<0$ to represent forces exerted by the actomyosin network in the cell cortex
which result in self-contraction. 
In addition, to account for the interactions between the cells, we include a contractile dipolar stress, $\zeta^\text{{a}}_2<0$, which acts between overlapping neighbours $j$ of ellipse $i$:
\begin{equation} 
\boldsymbol{\pi}^{n}_i=-\zeta^\text{{a}}_2 \sum_{ j \neq i}  \mathbf{d}_j { \varphi_j}(\mathbf{x}). \label{eq6}
\end{equation}
Self and neighbour active stresses are illustrated in Fig.~\ref{fig1main}(d),(e) for $\zeta_1^\text{a}<0$ and $\zeta_2^\text{a}<0$. 
They arise from cell-cell junctions and stress fibres exerting forces within and between cells, Fig.~\ref{fig1main}(f),(g)~\cite{eckert2023hexanematic,balasubramaniam2021investigating,choi2022cell,haas2022zo}.

The active force on the ellipse $i$, $\mathbf{f}^{\text{a}}_{i}(\mathbf{x}) = \boldsymbol{\nabla} \cdot (\boldsymbol{\pi}^s_i+\boldsymbol{\pi}^n_i)$, then reads \cite{PhysRevLett.130.038202}
 \begin{align}
\begin{split}
\mathbf{f}^{\text{a}}_{i}(\mathbf{x}) = -\zeta^\text{a}_1  \mathbf{d}_i \cdot {\boldsymbol{\nabla} \varphi_i}(\mathbf{x}) 
-\zeta^\text{{a}}_2 \sum_{ j \neq i}  \mathbf{d}_j \cdot {\boldsymbol{\nabla} \varphi_j}(\mathbf{x}) \\
 \approx -\zeta^\text{a}_1  \mathbf{d}_i \cdot {\boldsymbol{\nabla} \varphi_i}(\mathbf{x}) 
+ \zeta^\text{{a}}_2 \sum_{ j \neq i}  \mathbf{d}_{j} \cdot \boldsymbol{\nabla} \varphi_{i}(\mathbf{x}),
   \label{eq:active-force}
\end{split}
\end{align}
 where in the second step we have used the approximation $\boldsymbol{\nabla} \varphi_{j} \approx -\boldsymbol{\nabla} \varphi_{i}$ which is valid when the ellipses are close to confluent. 
Assuming overdamped dynamics and, for now, ignoring passive forces, the dynamics of the phase field is given by
 \begin{align}
\label{eq:intU}
 \dot{\varphi}_{i} = -  \frac{\textbf{f}_i^\text{a}}{\Gamma} \cdot \boldsymbol{\nabla} \varphi_{i} 
\end{align}
where $\Gamma$ is a friction coefficient, and an over-dot denotes a time derivative.
Using Eqs.~(\ref{eq:active-force}) and (\ref{eq:intU}), the Gaussian form (\ref{eq3}) chosen for the particle shape then allows an analytic calculation of the time evolution of the anisotropy tensor (\ref{eq:deformation_tensor}), leading to (see Sec.~I in the SM)
\begin{equation}
\dot{\mathbf{w}}_i  = - \left( \alpha_i \mathbf{w}_i+ 
\bar{\alpha}_i \mathbf{\tilde{w}}_i \right),
\label{shapedynamics}
\end{equation}
where $\mathbf{\tilde{w}}_i=\sum_{ j \neq i} \mathbf{w}_j$, $\bar{\alpha}_i=8\pi^2 \zeta^\text{{a}}_2/(27 \Gamma a_i)$ and $\alpha_i = -8\pi^2 \zeta^\text{{a}}_1/(27 \Gamma a_i)$.
It is also possible to calculate the effect of the active interactions on the dynamics of the ellipse area. Noting that
$a_i=\frac{1}{2} \int \varphi_{i} d\mathbf{x}$,
\begin{align}
\dot{a}_i &= \frac{1}{2}  \int d\mathbf{x} \dot{\varphi}_{i} 
= -   \int d\mathbf{x} \frac{\textbf{f}^a_{i}}{2 \Gamma} \cdot \boldsymbol{\nabla} \varphi_{i}  \nonumber \\ 
&=   c_i \boldsymbol{\textbf{w}}_i:\boldsymbol{\textbf{w}}_i + \bar{c}_i\boldsymbol{\textbf{w}}_i:
\mathbf{\tilde{w}}_i,
   \label{eq:Uan}
\end{align}
where 
$c_i=- \pi^4 \zeta_1^a/(32 a_i^2 \Gamma)$, $\bar{c}_i= \pi^4 \zeta_2^a/(32 a_i^2 \Gamma)$,
and the final step follows from Eq.~(\ref{eq:active-force}) and the definition of the anisotropy tensor~(\ref{eq:deformation_tensor}), and by assuming that the spatial variation of the area is slow compared to the cell anisotropy.

To introduce the active interactions into the hydrodynamic equations of the fields associated with particle shape and area we need to rewrite Eqs.~(\ref{shapedynamics})  and~(\ref{eq:Uan}) in terms of coarse-grained field variables defined as 
  \begin{align}
\textbf{W} & = \langle \textbf{w} \rangle = \int \textbf{w} \:\psi \:da \: d\textbf{w}\: d\textbf{r}, \label{eqs2m}\\
\mathcal{A} & = \langle a \rangle = \int a \:\psi \:da \: d\textbf{w}\: d\textbf{r} \label{eqs2mb}
\end{align}
where $\psi(\mathbf{r},t, a,\mathbf{w})$ is
the single particle probability distribution function.
The probability density function  evolves due to the particle velocity $\dot{\textbf{r}}$, anisotropy current $\dot{\mathbf{{w}}}$, and area current $\dot{a}$, and its dynamics reads
 \begin{align}
\partial_t \psi(\mathbf{r},t,a,\mathbf{d}) &= -\boldsymbol{\nabla}_r  \cdot (\dot{\textbf{r}} \psi) -\nabla_a  (\dot{a} \psi)-\boldsymbol{\nabla}_{\mathbf{ {w}}}  :(\dot{\mathbf{ {w}}} \psi).
   \label{eq:smolo}
   \end{align}

We use a mean-field approximation to write the effect of the neighbours in the dynamics in Eqs.~(\ref{shapedynamics}) and~(\ref{eq:Uan}) as
\begin{align}
&\dot{\mathbf{w}}_i  = - \left( \alpha_i \mathbf{w}_i+ 
\bar{\alpha}_i \bar{\textbf{W}} \right)\\
&\dot{a}_i = c_i \boldsymbol{\textbf{w}}_i:\boldsymbol{\textbf{w}}_i + \bar{c}_i \boldsymbol{\textbf{w}}_i:
\bar{\textbf{W}},
\label{shynamicsb}
\end{align}
where
$\bar{\textbf{W}} = \textbf{W} +\frac{ r_0^2 }{2}  \nabla^2 \textbf{W}$, $r_0$ is a length scale over which neighbouring ellipses interact, and we have assumed that the distribution function is isotropic in space.

Eqs.~(\ref{eqs2m})--(\ref{shynamicsb}) then give the contributions of internal and neighbour active nematic interactions to the dynamics of the fields as (see section II in the SM)
\begin{align}
 \dot{\textbf{W}}^I=&  -\bar{\alpha}  \bar{\textbf{W}}-\alpha \textbf{W},  \label{r5} 
\\ \dot{\mathcal{A}}^I=& \bar{c} \textbf{W} : \bar{\textbf{W}}+ \frac{c}{2} \textbf{W} :\textbf{W},  \label{r2}
\end{align} 
where $\bar{\alpha}=8\pi^2 \zeta^\text{{a}}_2/(27 \Gamma \mathcal{A}_0)$, $\alpha = -8\pi^2 \zeta^\text{{a}}_1/(27 \Gamma \mathcal{A}_0)$, $c=- \pi^4 \zeta_1^a/(32 \mathcal{A}_0^2 \Gamma)$, $\bar{c}= \pi^4 \zeta_2^a/(32 \mathcal{A}_0^2 \Gamma)$.\\

\noindent
\textit{Passive short-range interactions}—
\label{passiveshortrange}
We follow the standard practice of including thermodynamic interactions in the nematohydrodynamic equations for fixed-shape particles by incorporating a term  $ \caldot X^f =- \Gamma_{\mathcal{X}} \frac{\delta \mathcal{F}}{\delta \mathcal{X}}$ which describes relaxational dynamics to the minimum of a free energy $\mathcal{F}$ at a rate proportional to $\Gamma_{\mathcal{X}}$. 
The  free energy is defined as a Landau expansion in the order parameters, here the anisotropy tensor
 $\mathbf{W}$, and the area $\mathcal{A}$:
\begin{align}\label{ela}
\mathcal{F}&=
 \frac{C_\mathbf{W}}{4} (1+ \mathbf{W} : \mathbf{W})^2+\frac{K_\mathbf{W}}{2} |\nabla \mathbf{W}|^2 
 \nonumber \\ 
& + \frac{C_{\mathcal{A}}}{2} (\mathcal{A} -\mathcal{A}_0)^2 + \frac{C_{\mathcal{A}}^\prime}{4} (\mathcal{A} -\mathcal{A}_0)^4 +\frac{K_{\mathcal{A}}}{2} |\nabla \mathcal{A}|^2. 
\end{align} 

 In Eq.~(\ref{ela}) the term with coefficient $C_\mathbf{W}>0$ favours relaxation of the cells to an isotropic shape, and the term with the coefficient  $K_{\mathbf{W}}>0$ is an elastic free
energy density that tends to align the anisotropic nematogens.
The terms in area $\mathcal{A}$ favor relaxation to a homogeneous area $\mathcal{A}_0$ with the gradient term penalising inhomogeneities in area. The fourth-order terms in area and cell anisotropy are primarily included to aid numerical stability at large activities. \\

\noindent
\textit{Competition between active and passive interactions}—
\label{activeandpassive}
We next discuss how the active and passive interactions combine to affect the shape and relative orientation of the cells. From Eqs.~(\ref{r5}), (\ref{r2}) and (\ref{ela}) the equations of motion for particle shape and area are      
\begin{align}
\dot{\mathbf{W}}&= \left\{\Gamma_{\mathbf{W}} K_{\mathbf{W}}
 - \frac {\bar{\alpha}  r_0^2}{2}  \right\}  \nabla^2\mathbf{W} \nonumber \\
&-\{\bar{\alpha} +\alpha+\Gamma_{\mathbf{W}} C_{\mathbf{W}} (1+\mathbf{W}:\mathbf{W})\} \textbf{W},
\label{shortrangeaniso}\\
 \dot{\mathcal{A}}& =   -\Gamma_{\mathcal{A}} (C_{\mathcal{A}} (\mathcal{A}  -\mathcal{A}_0) +C_{\mathcal{A}}^\prime (\mathcal{A}-\mathcal{A}_0) ^3 - K_{\mathcal{A}} \nabla^2\mathcal{A})\nonumber \\ &
 +(\bar{c}+\frac{c}{2}) \textbf{W} :\textbf{W} + \frac{\bar{c} r_0^2}{2} \textbf{W} : \nabla^2 \textbf{W}. \label{shortrangearea}
\end{align}  
The first term in Eq.~(\ref{shortrangeaniso}) defines an effective elastic constant $\Gamma_{\mathbf{W}} K_{\mathbf{W}}
 - \bar{\alpha} r_0^2/2 $. 
 Note that this becomes negative for sufficiently large and positive active interactions between ellipses, destabilising any nematic alignment and suggesting that the nematogens prefer to align perpendicular to each other. Perpendicular alignment has been observed in mesoscopic, multi-phase field simulations of active deformable cell monolayers with  $\bar{\alpha}>0$ \cite{PhysRevLett.130.038202}, and in simulations of self-propelling hard rods \cite{mccandlish2012spontaneous}, and dividing cylindrical bacterial systems\cite{volfson2008biomechanical}. However, it means that our assumption of small deviation between neighbouring nematogens used in the mean field approximation and ignoring higher order moments of the distribution function is not correct and therefore we have restricted ourselves to considering $\bar{\alpha}<0$.
 
The second line of Eq.~(\ref{shortrangeaniso}) implies a competition between active and passive terms in changing the shape of the cells. 
Sufficiently negative $\alpha$ and $\bar{\alpha}$ destabilise the isotropic shape and lead to a homogeneous steady state solution for the elongation $|\mathbf{W}|_{st}$
given by
\begin{align}
&|\mathbf{W}|_{st}^2= -\frac{(\bar{\alpha}+\alpha)}{\Gamma_{\mathbf{W}} C_{\mathbf{W}}}-1. \label{steadyd}
\end{align} 
Fig.~\ref{fig1main}(e) shows schematically that contractile neighbour interactions with negative $\bar{\alpha}$ ($\zeta_2^a<0$, black arrows) favour elongation.

The homogeneous steady state area $\mathcal{A}_{st}$ then follows from Eq.~(\ref{shortrangearea}):
\begin{align}
 &\mathcal{A}_{st} =  \mathcal{A}_{0} +  \frac{(\bar{c}+ c/2)}{ \Gamma_{\mathbf{A}} C_{\mathbf{A}}} |\mathbf{W}|_{st}^2. \label{steadya}
\end{align}  
Eq.~(\ref{steadya}) indicates that the cell area decreases with elongation ($|\mathbf{W}|_{st}^2\neq 0$) in the presence of large contractile neighbour activity. 
The prediction agrees with multi-phase field simulations of active deformable cells \cite{PhysRevLett.130.038202} 
 which show hole formation in monolayers with  $\bar{c} \sim\zeta_2^a<0$ corresponding to a decrease in cell area. \\
 
     \begin{figure}[t] 
    \centering
    \includegraphics[width=0.5\textwidth]{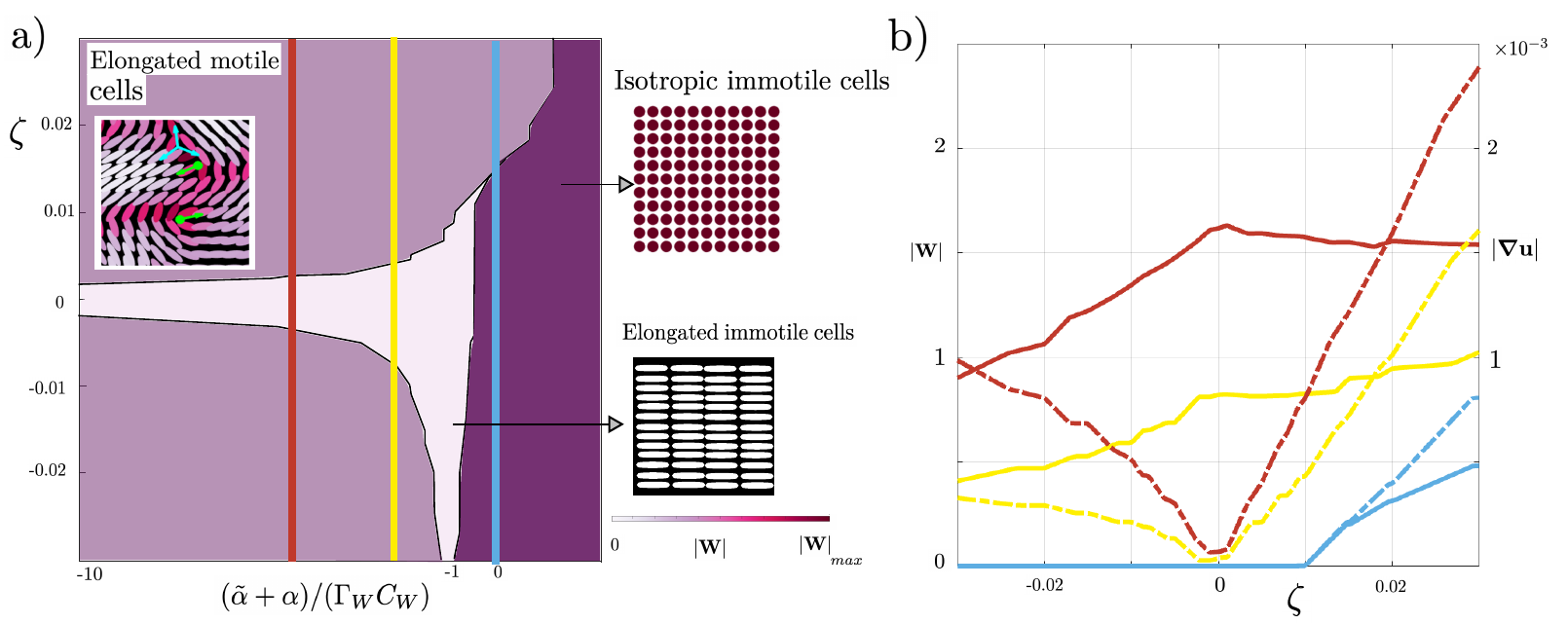}
    \caption{(a) Phase diagram for a deformable active monolayer. The $x$-axis describes short-range interactions (Eq.~(\ref{steadyd})) and the $y$-axis the effects of stresslet-induced flows 
    (Eq.~(\ref{nseq})). Three different phases are observed: a stationary isotropic phase, a stationary nematic phase and an active phase with non-zero velocity and defect formation. (b) Cell shape anisotropy (solid line, left axis) and velocity gradient (dashed line, right axis) along the lines marked in (a). The anistropic shape of the curves is because $\zeta>0$ ($\zeta<0$) facilitates (suppresses) cell elongation. The schematics show simulation results, where the aspect ratio is used to plot the ellipses.} 
  \label{fig2main}
\end{figure}

\noindent
\textit{Long-range flows}—
\label{flow}
In the SM, Section III, we show that the contribution of an imposed incompressible velocity field $\vec{u}$ to the dynamics of the anisotropy and area fields, assuming a small cell anisotropy $(\ell^{\parallel}_i/\ell^{\perp}_i)-1$, is
\begin{align}
\dot{\mathbf{W}}^u&=- \vec{u}\cdot \boldsymbol{\nabla} \mathbf{W} +\tens{\Omega}\cdot \mathbf{W} -\mathbf{W} \cdot \tens{\Omega}+ \tens{E}, \label{qnqe}\\
\dot{\mathcal{A}}^u &=  -\vec{u} \cdot \boldsymbol{\nabla} \mathcal{A} \label{r1}
\end{align}
where $\tens{\Omega}$ and $\tens{E}$ are the asymmetric and symmetric parts of the velocity gradient tensor.
The evolution of $\mathbf{W}$ is very similar to the dynamics of a nematic liquid crystal in an isotropic phase and with a positive flow aligning parameter $\lambda=1$~\cite{PRXLife.1.023008}.  The positive value of $\lambda$ shows the tendency of the isotropic particles to extend along the largest positive eigenvector of the strain rate tensor. 

As an example, we assume that the velocity field is created by the active particles themselves and that the dynamics of the incompressible fluid with density $\rho$ and viscosity $\eta$ is given by the Navier-Stokes equation,
\begin{eqnarray}
   & \rho (\dot{\vec{u}}+ \vec{u}\cdot \boldsymbol{\nabla} \textbf{u})= \boldsymbol{\nabla} P + \eta \nabla^2 \vec{u} +\zeta \boldsymbol{\nabla} \cdot \textbf{W},&\label{nseq} \\ & \boldsymbol{\nabla}\cdot \vec{u}=0,&\label{eqn:nsbb}
\end{eqnarray}
 where $P$ is the pressure field, and $\zeta \boldsymbol{\nabla} \cdot \textbf{W}$ is the usual activity term which results from coarse-graining the stresslet flow fields resulting from particle motion. 

Solving Eqs.~(\ref{eqmotion},\ref{nseq},\ref{eqn:nsbb}) numerically (see SM, Section IV) results in the phase diagram, and corresponding values for cell elongation and velocity gradients, shown in Fig.~\ref{fig2main}(a) and (b). Three different regions are observed. For 
$(\tilde{\alpha}+\alpha)/(\Gamma_W C_W)>-1$ (Eq.~(\ref{steadyd})) the cells remain isotropic and the velocity is zero unless $\zeta$ is sufficiently large and positive, when the extensional flows can elongate the particles to give a nematic state. As $(\tilde{\alpha}+\alpha)/(\Gamma_W C_W)$ decreases from $-1$, short-range forces between neighbours elongate the particles to give a nematic state, with hydrodynamic flows tending to enhance or suppress the elongation for $\zeta>0$ or $\zeta<0$, respectively. For sufficiently extended particles, there is an active instability and the nematic state becomes unstable, giving rise to active turbulence, characterised by a chaotic flow field and motile topological defects. \\

By coarse-graining a model of closely-packed ellipses that can vary their aspect ratio we have derived continuum equations for confluent deformable particles, an important example of which is epithelial cell layers. We show that contractile neighbour interactions, modelling the effects of actin filaments and stress fibers, compete with cell self contractility, and cause collective cell elongation and, when flows are permitted, form ±1/2 defects.
Our results reproduce many of the features seen in multi-phase field models which resolve the boundaries of individual cells including cell extension by neighbour interactions and changes in cell size.
The simplicity of the model suggests the possibility of studying how, for example, traction forces, active line tension,
 or different particle symmetries could be included in a continuum model of deformable particles.

\section*{Acknowledgements}
We thank Guanming Zhang, Delia Cropper, Ioannis Hadjifrangiskou, James Graham, and Sumesh Thampi for helpful discussions. JMY acknowledges support from the UK EPSRC
(Award EP/W023849/1) and ERC Advanced Grant ActBio (funded as UKRI Frontier Research Grant EP/Y033981/1).
\bibliographystyle{apsrev4-1}
\bibliography{references}

\end{document}